\documentclass[twocolumn,secnumarabic,amssymb, nobibnotes, prb]{revtex4}
 
\usepackage{bm}%
\usepackage{graphicx}
\usepackage{subfigure}
\usepackage{multirow}
\usepackage{color}

\usepackage{multirow}

\begin{document}
\author{A. Gr\"unebohm}%
\email{anna@thp.uni-due.de}
\address{Faculty of Physics and CENIDE, University of Duisburg-Essen, 47048 Duisburg, Germany}
\author{H. C. Herper}
\address{Department of Physics and Astronomy, Uppsala University,  751 20 Uppsala, Sweden}
\author{P. Entel}
\address{Faculty of Physics and CENIDE, University of Duisburg-Essen, 47048 Duisburg, Germany}

\title{On the rich magnetic phase diagram of (Ni, Co)-Mn-Sn Heusler alloys}
\begin{abstract}
 We put a spotlight on the exceptional magnetic properties of the metamagnetic Heusler alloy (Ni,Co)-Mn-Sn by means of first principles simulations. 
 In the energy landscape we find a multitude of local minima, which belong to different ferrimagnetic states and are close in total magnetization and energy.
All these magnetic states correspond to the local high spin state of the Mn atoms with different spin alignments and are related to the  magnetic properties of Mn. Compared to pure Mn, the magneto-volume coupling is reduced  by Ni, Co, and Sn atoms in the lattice and no local low-spin Mn states appear. 
For the cubic phase we find a ferromagnetic ground state whereas the global energy minimum is a tetragonal  state with complicated spin structure and vanishing magnetization which so far has been overlooked in simulations.
\end{abstract}
\maketitle
The giant magnetocaloric effect (MCE) has attracted huge attention being viewed as environment friendly alternative for common cooling devices especially after the  observation of this effect at room temperature in 4{\em f} alloys, such as Ga$_5$Si$_{4-x}$Ge$_x$ \cite{Pecharsky, Gschneidner}. This caloric effect  is related to a first order magneto-structural phase transition at the martensitic transition temperature $T_M$.
Below $T_M$ a small external magnetic field can induce the metamagnetic transition, a feature which allows for giant adiabatic temperature changes. But 4f alloys also bring their own environmental problems into the picture and today the search  rare earth free alternatives working at ambient conditions is in full swing.  Recently Mn-rich Ni$_2$-Mn based Heusler alloys   especially, Ni-Mn-In, Ni-Mn-Ga, and Ni-Mn-Sn, became the focus of attention \cite{Liu2,Planes,Buchelnikov,Sokolovskiy2,Graf,Klaer,Fabbrici}.
In particular Ni-Mn-Sn alloys  with  large refrigeration capacity at ambient temperatures  are promising for applications \cite{Huang}.
In all these Ni-Mn based alloys, a subtle interplay between ferromagnetic and antiferromagnetic interactions has been observed.
With increasing Mn concentration, the antiferromagnetic interactions increase resulting in a reduced magnetization, especially in the low temperature phase. At the same time $T_M$ increases systematically \cite{NiMnIn3,Graf,Krenke}.

  In addition, the substitution of small amounts of Ni by Co can be used to optimize the effect as  magnetic and structural transition temperatures and magnetization are modified  \cite{Sokolovskiy2,Sokolovskiy4,Cong, Intermag14,Fabbrici}.
So far the application is handicapped by the irreversibility of the  caloric response  \cite{Umetsu,Ghosh,Titov,Huang,Bhatti} and a fundamental understanding of the magnetic phases and their impact on the metamagnetic transition is urgently needed for further optimization. 
The magnetic properties of Ni(Co)-Mn-(Sn,Ga,In)  are far from being understood; paramagnetic, ferromagnetic, anti-ferromagnetic, non-collinear, spin-glass phases,  FM clusters in an AF matrix, exchange bias phenomena, and reoccurrence transitions, all have been found -- depending on temperature, composition, field treatment, and sample preparation \cite{Sokolovskiy, Cong, Das, Kanomata, Jing, Krenke,Krenke2,Umetsu,Chen,Intermag14,Ye,Krumme,Cakir,Ghosh,Bhatti,Yuan,Cakir2,Graf}. We concentrate on Ni$_{43.75}$Co$_{6.25}$Mn$_{43.75}$Sn$_{6.25}$ (Ni$_7$CoMn$_7$Sn). 
Our interests are the stability of different magnetic structures and the first order transition from antiferromagnetic like spin alignment at low temperature to ferromagnetic order at high temperatures as in Mn$_3$Ga \cite{Cakir3} or FeRh \cite{Kouvel}. In addition, the  long-range magnetic order maybe be broken in Ni$_7$CoMn$_7$Sn due to the large atomic disorder, and competing FM and AF trends may be possible at all temperatures.

In order to shed light on the magnetic structure of the alloy we make use of the fixed spin moment approach and constrain the overall magnetic moment of the simulation cell.
This approach has been used successfully to sample the different magnetic phases of 3d elements \cite{Haeglund, Moruzzi,Duschanek} and to discuss the influence of an external magnetic field in Ni$_2$MnIn \cite{Entel4}.
Our inspection of the energy landscape as a function of magnetization reveals a local energy minimum for each state with all Mn atoms being in their high-spin state. The local minima for different relative orientations of the Mn spins are close in energy and magnetization. For realistic system sizes, the large number of minima may allow for thermally activated spin-flip transitions in the cubic material. For the tetragonal low temperature phase of Ni$_7$CoMn$_7$Sn we find an antiferromagnetic ground state  which has not been taken into account previously.

The paper is organized as follows:  Technical details are  given in Sec.~\ref{sec:method} and the Appendix. Our results for the structural and magnetic properties are presented in Sec.~\ref{sec:mag} and the energy landscape for different values of the magnetization for pure Mn and Ni$_7$CoMn$_7$Sn is discussed in Sec.~\ref{sec:FSM}. Conclusions and outlook can be found in Sec.~\ref{sec:conclude}.
\section{Technical details}
\label{sec:method}
\begin{figure}
\centering
\includegraphics[width=0.3\textwidth,clip,trim=2.5cm 3cm 2.5cm 3cm]{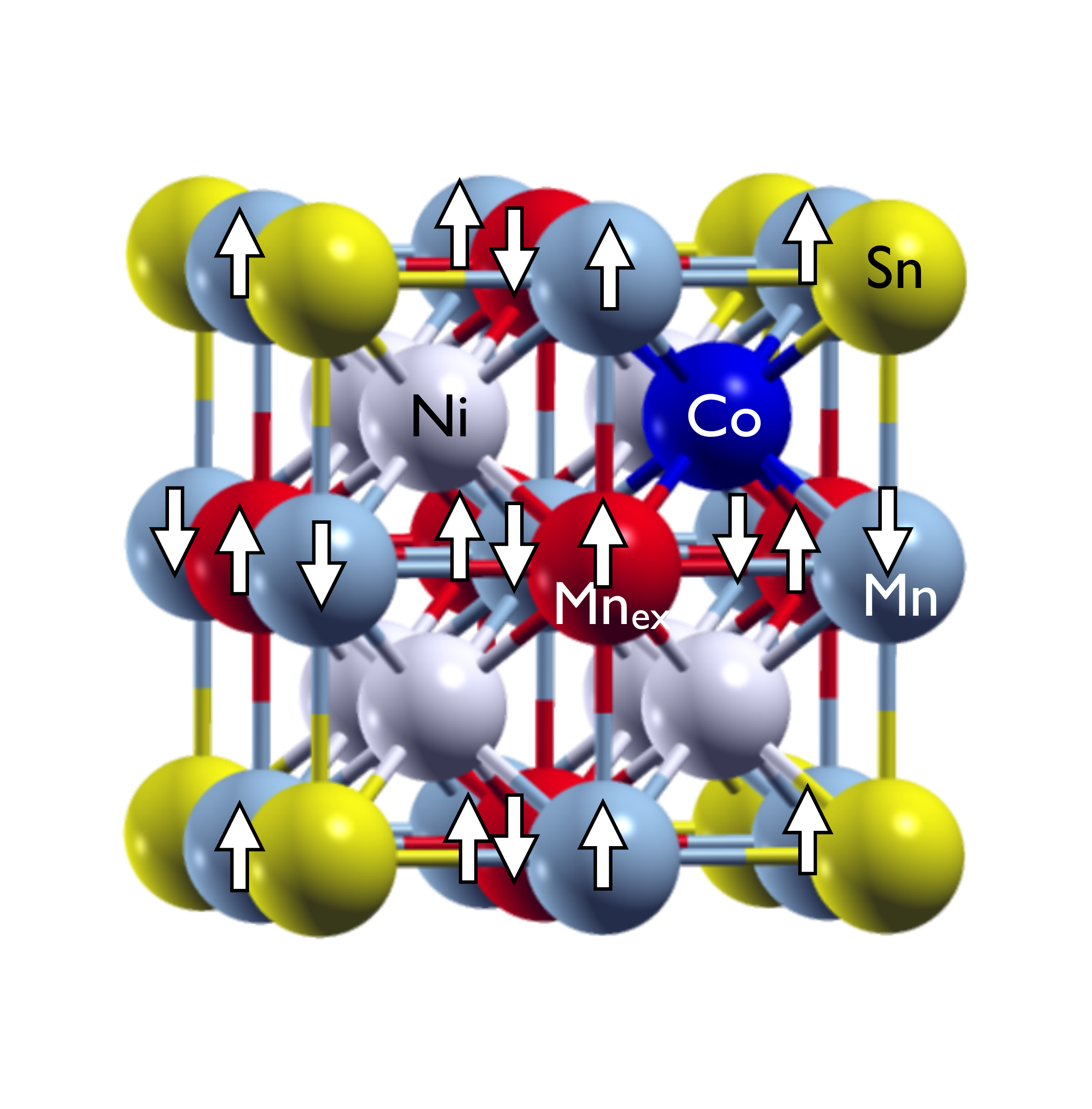}
\caption{Structure and atomic distribution of Ni$_7$CoMn$_7$Sn in the supercell. Different colors and gray shades are chosen to represent  Sn, Ni, Mn,and Mn excess atoms on the Sn lattice, respectively. Arrows illustrate the spins of Mn atoms for the configuration mix$_{{1}}$.\label{fig:struct} }
\end{figure}

The magnetic and structural  properties of Ni$_7$CoMn$_7$Sn have been calculated with the plane-wave density functional theory code VASP \cite{Kresse3}. Projector augmented wave potentials in the PBE approximation \cite{PBE} have been utilized treating 13, 9, 16, and 14 electrons as valence for Mn, Co, Ni, and Sn, respectively.
The ionic positions have been optimized  until forces have converged to an accuracy of 0.01~eV/{\AA}.  Unless otherwise stated, we have sampled the energy landscape for the fixed lattice constant of the established ferrimagnetic phase (ferri$_1$:~5.87~{\AA}) and used a supercell containing 16 atoms, cf.\ Fig~\ref{fig:struct}.  
 A plane-wave cutoff of 337.3 eV, a $8\times8\times8$ $k$-point grid  constructed with the Monkhorst-Pack scheme \cite{Monkhorst},  and convergence of the total energy to 10$^{-6}$~eV guarantee high accuracy of our results.
  In order to improve the convergence with respect to the $k$-mesh further, a Methfessel-Paxton smearing of 0.1~eV has been used. 

We have applied the fixed spin moment approach \cite{fsm} as implemented in the VASP code, i.e.\ the relation between majority and minority spins has been fixed to stabilize certain values of M$_{tot}$. As non-collinear coupling between an external field and magnetic moments, as well as spin-flip transitions, are not accessible with this approach, no direct link exists
between an external magnetic field and this imposed constraint.
Furthermore we have  fixed $M_{tot}$, neglecting different possible staggered magnetizations in the anti- or ferrimagnetic states. Thus in principle, we obtain the most favorable staggered magnetization for each $M_{tot}$. Due to the finite energy barriers for transitions  between magnetic phases with different symmetries, 
which are connected by spin flip transitions, we can account for these metastable states in our  simulations at zero temperature. 
These different spin states persist if we release the constraints on the magnetization.  

\begin{table*}
\centering
\caption{Energy difference relative to the cubic FM phase ($\Delta E$) and magnetic moments (M$_{tot}$) of different spin configurations of Ni$_7$CoMn$_7$Sn. 
The relativ numbers of Mn$\uparrow$ to Mn$\downarrow$ spins change systematically from 4:3 for configurations (1) to 7:0 in the full FM configuration.
 Mn$\downarrow$ spins are distributed on Mn$_{ex}$ sites, Mn sites, or on both classes of Mn positions for ferri$_i$, Mn$_i$, and mixed$_i$ states, respectively.
 The ground state lattice constant of the ferri$_1$ phase at $c/a=1$  has been used (partly results for an optimized lattice constant are included in brackets).
The supercell has been doubled along a for configurations (*). 
 \label{tab:free}}
\begin{tabular}{lcccccc}
\hline
\hline
&&\multicolumn{2}{c}{c/a=1}&\multicolumn{2}{c}{c/a=1.28}\\
\multicolumn{2}{c}{Magnetic}& $\Delta E$ &M$_{tot}$ &$\Delta E$ &M$_{tot}$\\
\multicolumn{2}{c}{configurations} &  (meV/f.u.)&$(\mu_B$/f.u.)&(meV/f.u.)&($\mu_B$/f.u.)\\
\hline
\multirow{2}{*}{(1): 4$^{\uparrow}$3$^{\downarrow}$}&ferri$_1$& 30 (30)&1.0 (1.0)&-89~~~~~~~~&1.0~~~~~~~~\\ 
&mix$_1$&88 (88)&1.2 (1.2)&{\bf{-225}} (-233)&1.2 (1.2)\\ 
\hline
\multirow{3}{*}{(2): 5$^{\uparrow}$2$^{\downarrow}$}&ferri$_2$&32 (32)&3.4 (3.4)&-39~~~~~~~~&3.2~~~~~~~~\\ 
&Mn$_2$&58 (57)&3.5 (3.5)&-28~~~~~~~~&3.3~~~~~~~~\\ 
&mix$_2$&79 (78)&3.5 (3.5)&-104~~~~~~~~&3.4~~~~~~~~\\ 
\hline
(*):  11$^{\uparrow}$3$^{\downarrow}$&mix$_{2.5}$&\multicolumn{2}{c}{--}&-31~~~~~~~~&4.4~~~~~~~~\\
\hline
\multirow{2}{*}{(3): 6$^{\uparrow}$1$^{\downarrow}$}&ferri$_3$&23 (21)&5.5 (5.6)&42~~~~~~~~&5.1~~~~~~~~\\ 
&Mn$_3$&78 (75)&5.7~~~~~~~~&119~~~~~~~~&5.3~~~~~~~~\\  
(*): 12$^{\uparrow}$2$^{\downarrow}$&mix$_3$&\multicolumn{2}{c}{--}&75~~~~~~~~&5.3~~~~~~~~\\ 
\hline
(*): 13$^{\uparrow}$1$^{\downarrow}$&ferri$_4$ &8~~~~~~~~&6.6~~~~~~~~&\multicolumn{2}{c}{--}\\
\hline
(FM): 7$^{\uparrow}$&FM&{\bf{0/(-6)}}&7.7 (7.8) &136 (134)&7.4 (7.4) \\
\hline
\hline
\end{tabular}
\end{table*}
Additionally, we have used the AkaiKKR package \cite{Ogura} in order to determine the pairwise magnetic exchange parameters employing Liechtenstein's formula \cite{Lichtenstein} in the framework of the KKR approach.
Muffin-tin potentials and the GGA PW91 approximation \cite{PW91} together with a maximal angular momentum of 3 have been used.
 For the self-consistence iteration of the cubic or tetragonal cell, and the calculation of the  magnetic exchange interactions, 29, 59, and 512  k-points have been taken into account. 
For this part of the calculations atomic relaxations have been neglected and the lattice constants as obtained by the supercell calculations have been used. While the supercell simulations are based on one chosen distribution of excess Mn and Co atoms, cf.\ appendix, the coherent potential approximation, as used in the KKR simulations, allows for a uniform distribution of atoms on each sublattice.
Despite the fact that the two methods applied in this study  differ in the accuracy of the used potentials  (muffin tin vs.\ PAW) and the description of disorder (CPA  without short-range interactions vs.\ local disorder in a super cell), both approaches yield comparable magnetic phases and magnetic moments, e.g., magnetic moments of  1.0~$\mu_B$/f.u. and 7.7~$\mu_B$/f.u. have been consistently found for the ferri$_1$ and FM states, respectively. Analogous it has been shown for several Heusler alloys in Ref.~\cite{Oedogan} that the local interactions, which are neglected in CPA, have little influence on the electronic and magnetic structure. 

The pairwise magnetic exchange interactions up to the distance of 3 lattice constants have been used as input for a classical Heisenberg model in order to determine the magnetization at finite temperatures ($M(T)$) by means of Monte Carlo simulations, cf.\ Ref.~\cite{Intermag14}.
Here, $M(T)$ has been modeled in cooling simulations for both structural phases separately within a simulation cell containing at least 3456 atoms with random distribution of Mn$_{ex}$ and Co atoms. 
\section{Magnetic and structural properties }
\label{sec:mag}
Ni$_7$CoMn$_7$Sn  crystallizes in the L2$_1$ structure, which is composed of four interpenetrating face centered cubic lattices, see Fig.~\ref{fig:struct}. It is known from literature that Co atoms mainly occupy Ni sites whereas the additional Mn$_{ex}$ atoms occupy Sn sites \cite{Ye, Kanomata, Jing}.
Under cooling, the cubic structure is destabilized at the martensitic transition temperature (T$_M$).
A systematic experimental investigation of the structural properties of Ni-Mn-Sn in dependence of Mn concentration and temperature has been performed in \cite{Cakir}.
For high temperatures and well below $T_M$, only the non-modulated L2$_1$ (cubic), and L1$_{0}$ (tetragonal, $c/a>1$) phases have been observed. 
It has been found that T$_M$ increases systematically with the Mn excess atoms similar to Ni-Mn-Ga and Ni-Mn-In \cite{Graf}. 
For at least 8.4 valence electrons per atom only  L2$_1$  and L1$_{0}$ have been found whereas different modulated phases exit  for lower Mn concentration. 
In particular, the  7M phase has been found as transition state between 
  L2$_1$, and L1$_{0}$ for  the chosen composition (8.24 valence electrons per atom). The 7M phase corresponds to 
  a modulated tetragonal phase with $c/a<1$ and superimposed monoclinic distortions. Alternatively this phase can be interpreted as an adaptive structure composed of  L1$_{0}$ nano twins \cite{Gruner,Kaufmann}.

A previous investigation of Ni$_7$CoMn$_7$Sn has been restricted to the FM state and the state with anti-ferromagnetic excess Mn spins (ferri$_1$) \cite{Intermag14}.
Interestingly, the stability of both phases changes with the tetragonal distortion.  The FM state is most favorable for $c/a=1$,  whereas the ferri$_1$ state is favored in the tetragonal state ($c/a=1.28$). Similarly, weakly magnetic and ferromagnetic tetragonal and cubic phase have been found, e.g., in Ni$_{48}$Co$_{5}$Mn$_{35}$In$_{12}$ \cite{Klaer}

\begin{figure}
\centerline{
\includegraphics[height=0.35\textwidth,clip, trim=0cm .5cm 0.5cm .5cm]{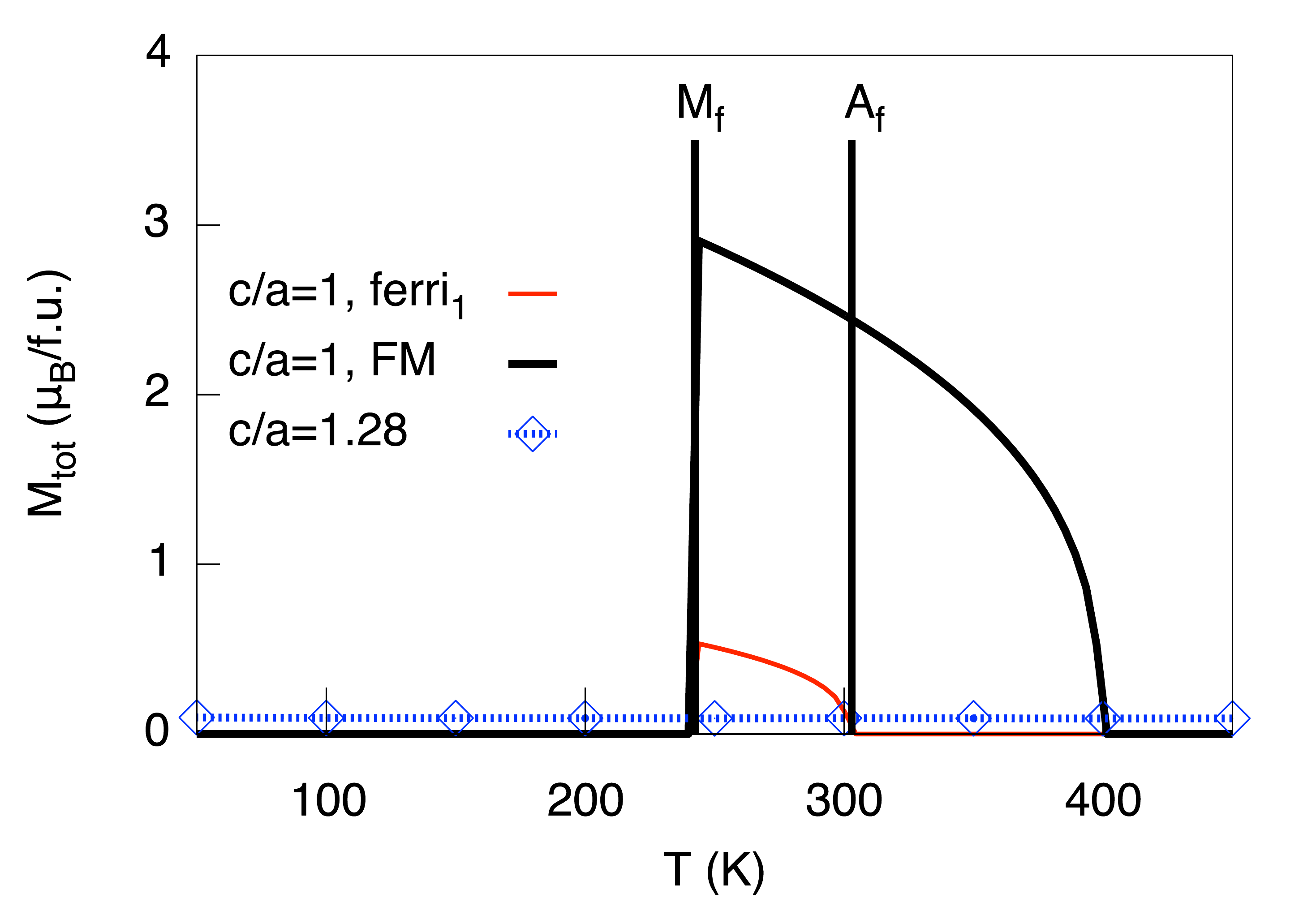}}
\caption{
Magnetization of Ni$_7$CoMn$_7$Sn at finite temperatures fitted by $M^{1/\beta}(T)$ to  reduced finite size effects, with $\beta$ the critical exponent of the Heisenberg model, cf.~\cite{Intermag14}. Black solid line: cubic phase (FM reference); Red dotted line: cubic phase (ferri$_1$ reference); Blue line with symbols: tetragonal phase (ferri$_1$ reference).
Vertical lines illustrate the experimental transition range of the structural transition of Ni$_{7.2}$Co$_{0.48}$Mn$_{6.9}$Sn$_{1.1}$ taken from Ref.~\cite{Chen}. 
\label{fig:jij_nimnsn}}
\end{figure}

In the current paper we have extracted the pair-wise magnetic exchange interactions for both these configurations and used them as input for a classical Heisenberg model.
Figure~\ref{fig:jij_nimnsn} shows the resulting temperature dependent magnetization. For the cubic phase we find an ordered magnetic state below the transition temperatures of $T_C\sim 300$~K and $T_C\sim 400$~K for the ferri$_1$ and FM reference state, respectively.
In qualitative agreement with non-collinear {\it ab initio} simulations, we find a canting of the Mn spins in our MC simulations, resulting in a reduction of the magnetization compared to the FM collinear state, compare Tab.~\ref{tab:free}.
For the  tetragonal phase ($c/a$=1.28) we find an antiferromagnetic phase (mix$_1$) without overall magnetization in the whole sampled temperature range (5--800 K), in agreement with the long-rage AF state found experimentally for the martensitic phases for Mn-rich Ni-(Co)-Mn-Sn alloys \cite{Bhatti,Behler,Pramanick}.
We note that in experiment non-ergodic phenomena and exchange bias coupling in these samples could be related to embedded FM clusters.
Such clusters are not accessible for the chosen size of the simulation cell and the random but uniform distribution of Co and Mn atoms.

\begin{figure*}
\begin{center}
\includegraphics[width=\textwidth]{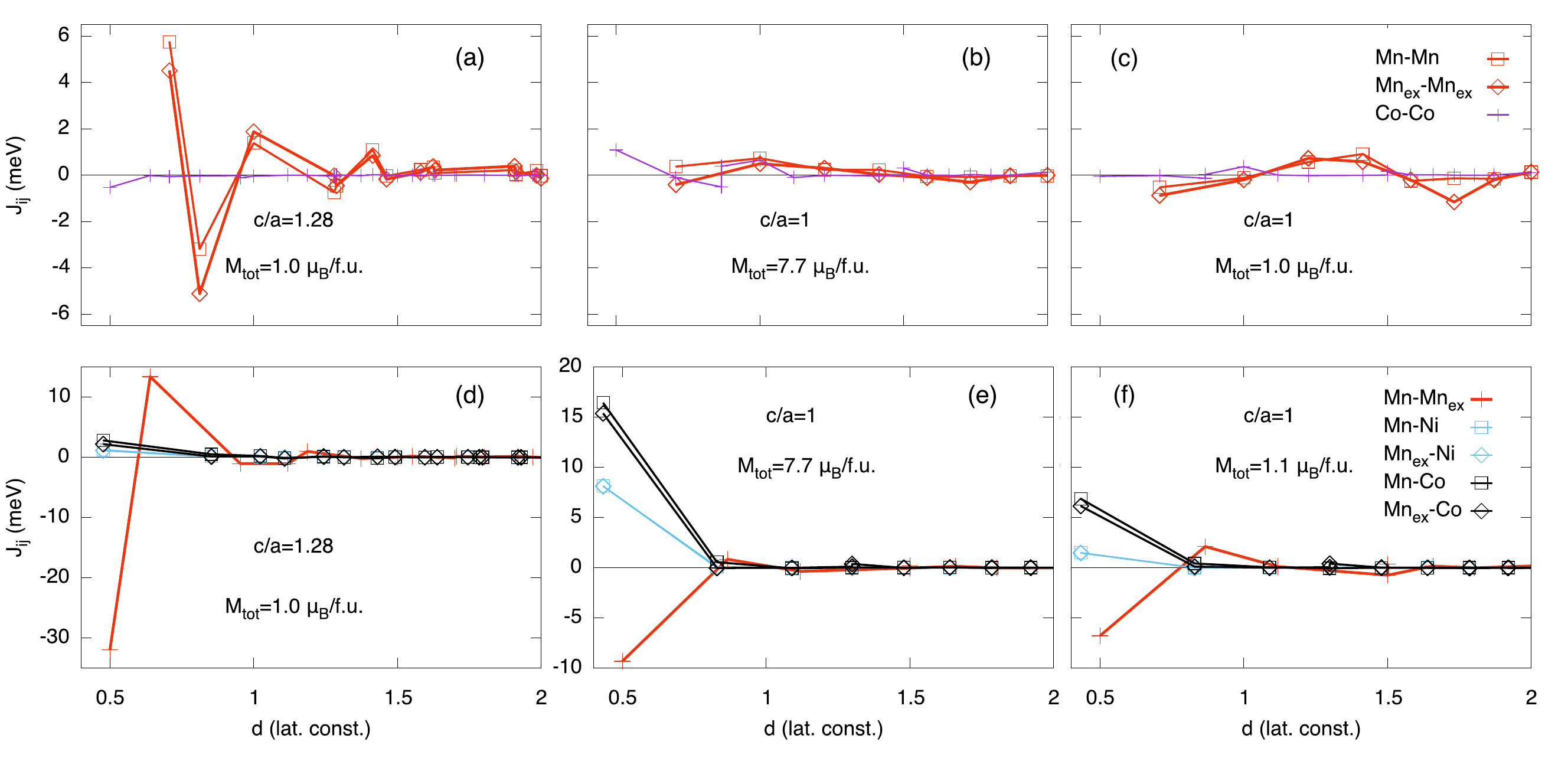}
\caption{Distance dependent magnetic exchange interactions in Ni$_7$CoMn$_7$Sn for different pairs of atoms in the lattice. Top: Intra-sublattice interaction; Bottom: inter-sublattice interactions (a) and (d) $c/a=1.28$; Reference state ferri$_1$, i.e.\  
antiparallel Mn$_{ex}$ spins;
(b)-(c) and (e)-(f) Cubic phase (b) and (e) ferri$_1$ reference state; (c) and (f) FM reference state;  \label{fig:Jij}  Mn$_{ex}$: Mn excess atoms on Sn lattice.}
\end{center}
\end{figure*}

The exceptional modifications of the magnetic phase diagram with strain and the magnetic reference configurations can be understood in terms of the frustrated magnetic interactions, see Fig.~\ref{fig:Jij}.
In all cases, localized ferromagnetic Mn-(Ni/Co) interactions compete with Mn-Mn interactions which oscillate between AF and FM with the atomic distances.
In full agreement localized FM Mn-Ni and Mn-Co inter-sublattice and oscillating intra-sublattice Mn-Mn interactions have been found in various Mn based Heusler alloys \cite{Sasioglu,Kurtulus,Rusz,Meinert2}.
The latter could be related to Ruderman-Kittel-Kasuya-Yoshida (RKKY) and super-exchange interactions mediated by $sp$ electrons at the Fermi level \cite{Rusz,Sasioglu3}.
It should be mentioned that the nearest Mn-Mn$_{ex}$ distance is significantly smaller than the intra-sublattice Mn-Mn distances such that a direct AFM exchange is more likely in this case.

Within the cubic phase, the FM couplings between nearest neighbors Mn and Ni/Co spins dominate the magnetic exchange interactions and the effective magnetic interactions for distances up to one lattice constant is ferromagnetic.

For the cubic structure, the ferri$_1$ state is  only 30~meV/f.u.\ less favorable than the FM  state.
Figure~\ref{fig:Jij}~(c) illustrate the influence of this reference spin alignment on the magnetic interactions.
 All magnetic interactions are reduced and the Mn-Mn$_{ex}$ interactions show an additional FM contribution for the next-nearest atoms with a distance of about 5.08~{\AA}.   

In the FM state,  Mn-Co and Mn-Ni spins are aligned nearly parallel and flipping the spin is costly due to the strong  FM  interactions, resulting in a high $T_C$.
Less energy is needed to turn the spins, resulting in a lower $T_C$, for the  ferri$_1$ reference state as the effective FM interactions are reduced by a factor of about ten, cf.\ Fig.~\ref{fig:Jij}~(b)--(c).

Tetragonal distortion ($c/a>1$) modifies the  Mn-Mn$_{ex}$ distances and the AF interactions between Mn and Mn$_{ex}$ atoms increase, destabilizing the  FM spin alignment \cite{Entel5}.
First, there is the strong direct AF interaction between Mn$_{{ex}}$ and Mn. In addition, the Mn-Mn distances along c/in c planes  increase/decrease inducing additional FM/AF interactions, which stabilize the mix$_1$ phase.
The  approximate collinear spin configuration of this phase is illustrated by arrows in Fig.~\ref{fig:struct}. In each $c$ layer,  
Mn-Mn and  Mn$_{ex}$-Mn$_{ex}$ spins are aligned parallel, whereas the Mn spins along $c$ and the Mn and Mn$_{ex}$ spins in each $c$ layer are aligned antiparallel.
Subsequent  {\it{ab initio}} simulations reveal that the mix$_1$ state, which has been overlooked so far, is the most favorable state also for the supercell simulations  at $T=0$~K, see Tab.~\ref{tab:free}.  
In the Monte Carlo simulation we use at least 1728 Mn spins and allow for non-collinear spin alignments resulting in a vanishing magnetization. 
In our {\it{ab initio}} simulations we use a supercell with 16 atoms  resulting in one uncompensated Mn spin per cell. In addition, we do not allow for the  non-collinear alignment of Ni and Co  spins resulting in spin moments of 0.1~$\mu_B$ (0.5~$\mu_B$) for each Ni (Co) atom. 
We obtain a magnetic moment of 1.2~$\mu_B$/f.u. for the approximated mix$_1$ state.

\section{Phase diagram under constrained magnetization}
\label{sec:FSM}
The magnetism of Ni$_7$CoMn$_7$Sn is dominated by Mn atoms which carry the largest moments and contribute most to the magnetic exchange interactions, whereas the magnetic characteristics of Ni and Co are mainly induced by the Mn environment. 
For instance, the Mn-Co interactions increase from about 6 meV to about 15~meV between ferri- and FM Mn neighborhood.
In addition, Ni (Co) moments vary between $\leq$0.1~$\mu_B$ (0.8~$\mu_B$/atom)  and 0.6~$\mu_B$ (1.5~$\mu_B$/atom) in the ferri$_1$ and FM phases, respectively.
The same dominance of Mn has been found in various Mn-based Heusler alloys, e.g.\ in Refs.~\cite{Buchelnikov,Sasioglu3}. 
  Due to the central role of Mn for the magnetism in this  compound we have in a first step studied an artificial compound placing Mn on all four sublattices of the Heusler system. 
\subsection{First insight: The pure Mn system}
\begin{figure}
\centering
\includegraphics[width=0.48\textwidth,clip, trim=3cm 3cm 2cm 2.7cm]{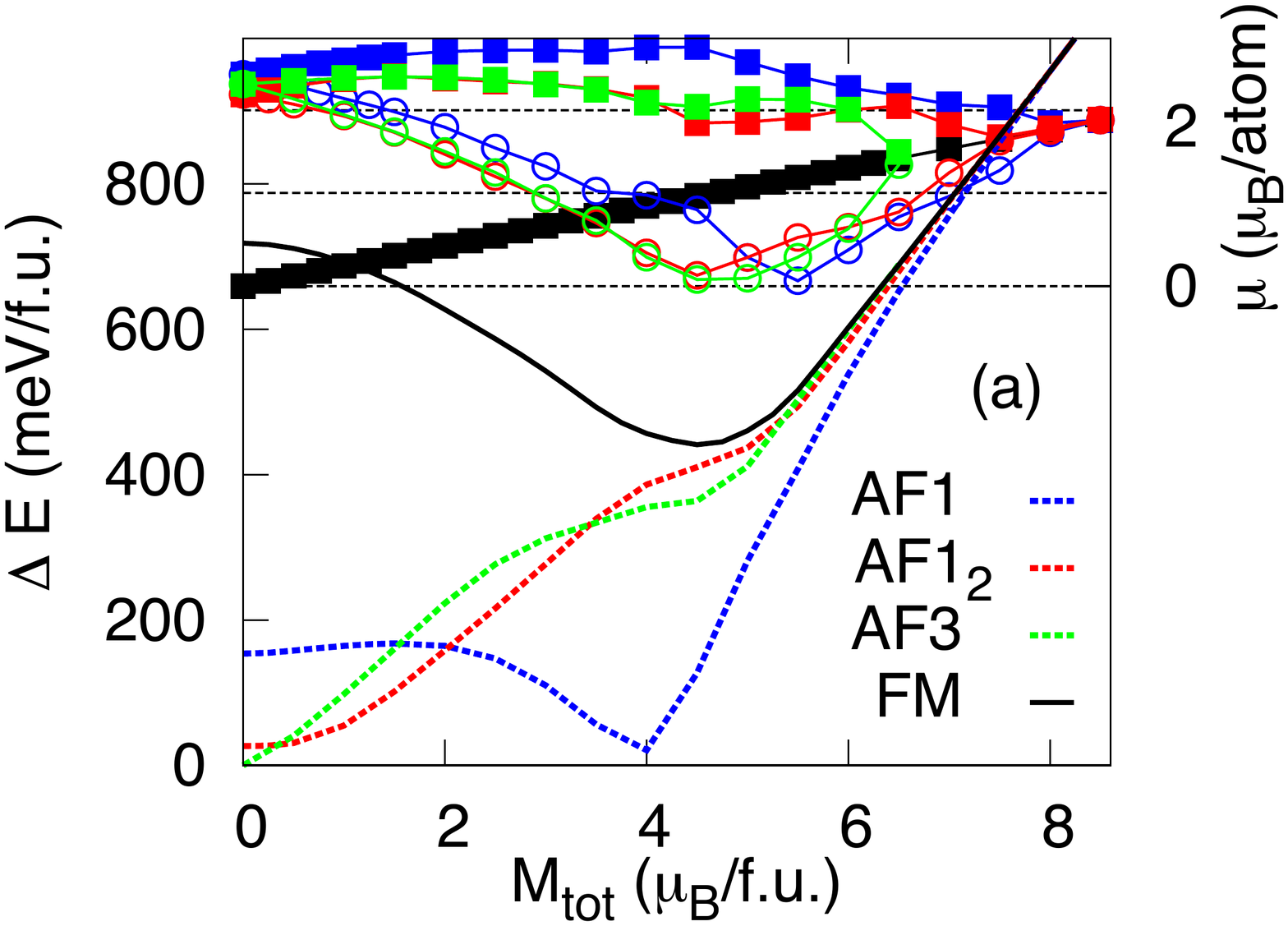}\\
\includegraphics[width=0.48\textwidth,clip,trim=0cm 0.6cm 0cm .3cm]{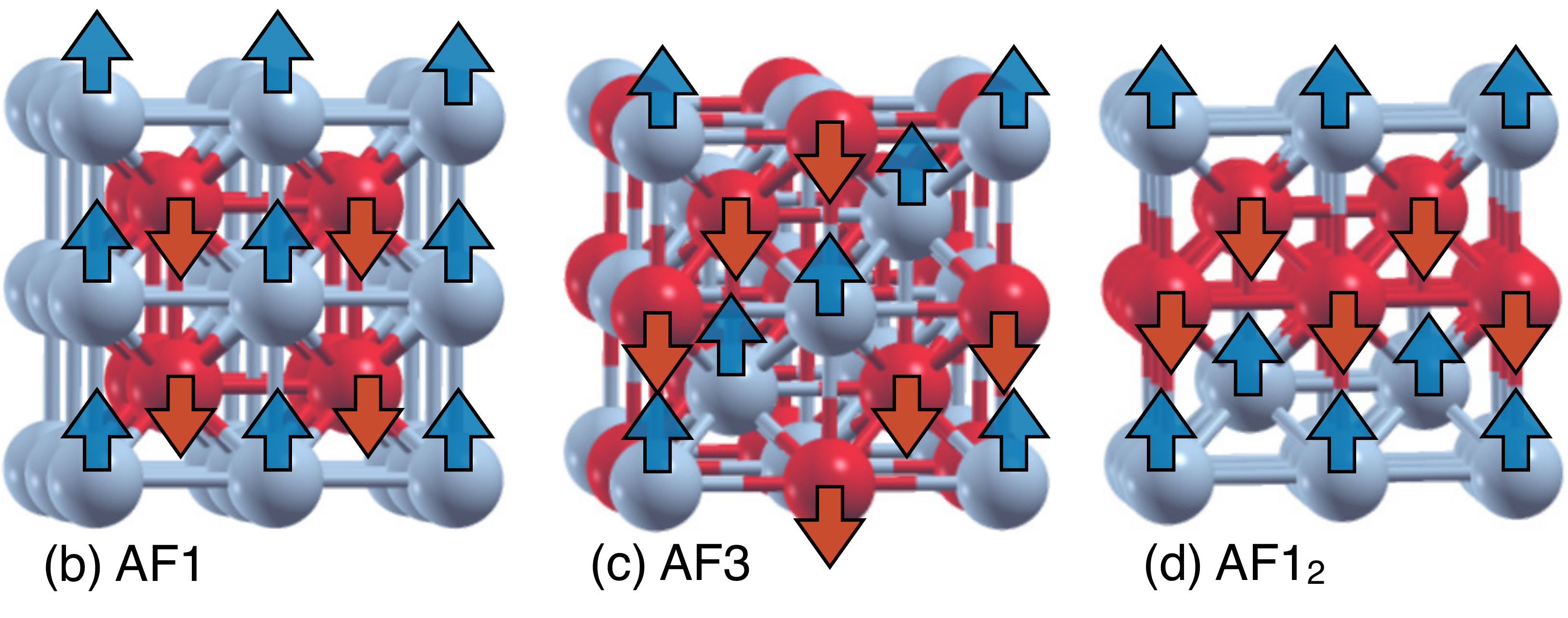}
\caption{(a) Total energy of pure Mn constrained to the lattice of Ni$_7$CoMn$_7$Sn for the spin configurations illustrated in (b)-(d).
The energy relative to the AF1 phase is given on the left axis (lines without symbols) and corresponding magnetic moments are given on the right axis (lines with symbols). Filled squares Mn$\uparrow$, open circles Mn$_{\downarrow}$.
 (b)-(d) Corresponding spin configurations. Arrows and color encode the direction of the spins.
\label{fig:EM}}
\end{figure}
In the five allotropic phases of Mn, spin moments of about 3.5 $\mu_B$/atom (high-spin),  1--2~$\mu_B$/atom (low-spin), and 0~$\mu_B$ (quenched)  exist \cite{Duschanek,Sliwko, Hafner}.
In the following the terms high-spin and low-spin are used for the corresponding local Mn moments.
 
Figure \ref{fig:EM} shows the energy landscape of pure Mn constrained to the cubic lattice structure and lattice constant of Ni$_7$CoMn$_7$Sn  for different spin alignments.
We adopt the nomenclature used for Fe-Rh \cite{Uebayashi} for the different states: AF1 for AF coupling between adjacent layers, AF3 for AF coupling along all three cartesian directions, and AF1$_2$ for AF double layers, see Fig.~\ref{fig:EM}~(b)-(d).
For all AF states the Mn atoms are in the high-spin state for M$_{tot}=0$ with magnetic moments between 2.2--3.1~$\mu_B$/atom.
With increasing $M_{tot}$, the (Mn$_{\downarrow}$) moments (open circles in Fig.~\ref{fig:EM}) are first quenched, reverse their sign, and finally the FM state is reached.
Similar to Fe-Rh, the most favorable state is AF3 with M$_{tot}=0$. For the configuration AF1, we also find a local energy minimum at $M_{tot}=4\mu_B$/f.u. and $\Delta E=21$~meV/f.u, with 
 alternating high- and low-spin Mn spins. This configuration turns out most favorable  in KKR-CPA simulations without constraints on $M_{tot}$. If all Mn spins are aligned parallel, only one  local energy minimum is found, the FM low-spin state at 4.4~$\mu_B$/f.u. and $\Delta E=440$~meV/f.u., which is also confirmed by 
KKR-CPA simulations.
\begin{figure}
\centerline{
\includegraphics[height=0.35\textwidth,clip,trim=4.5cm 17cm 3.6cm 4cm]{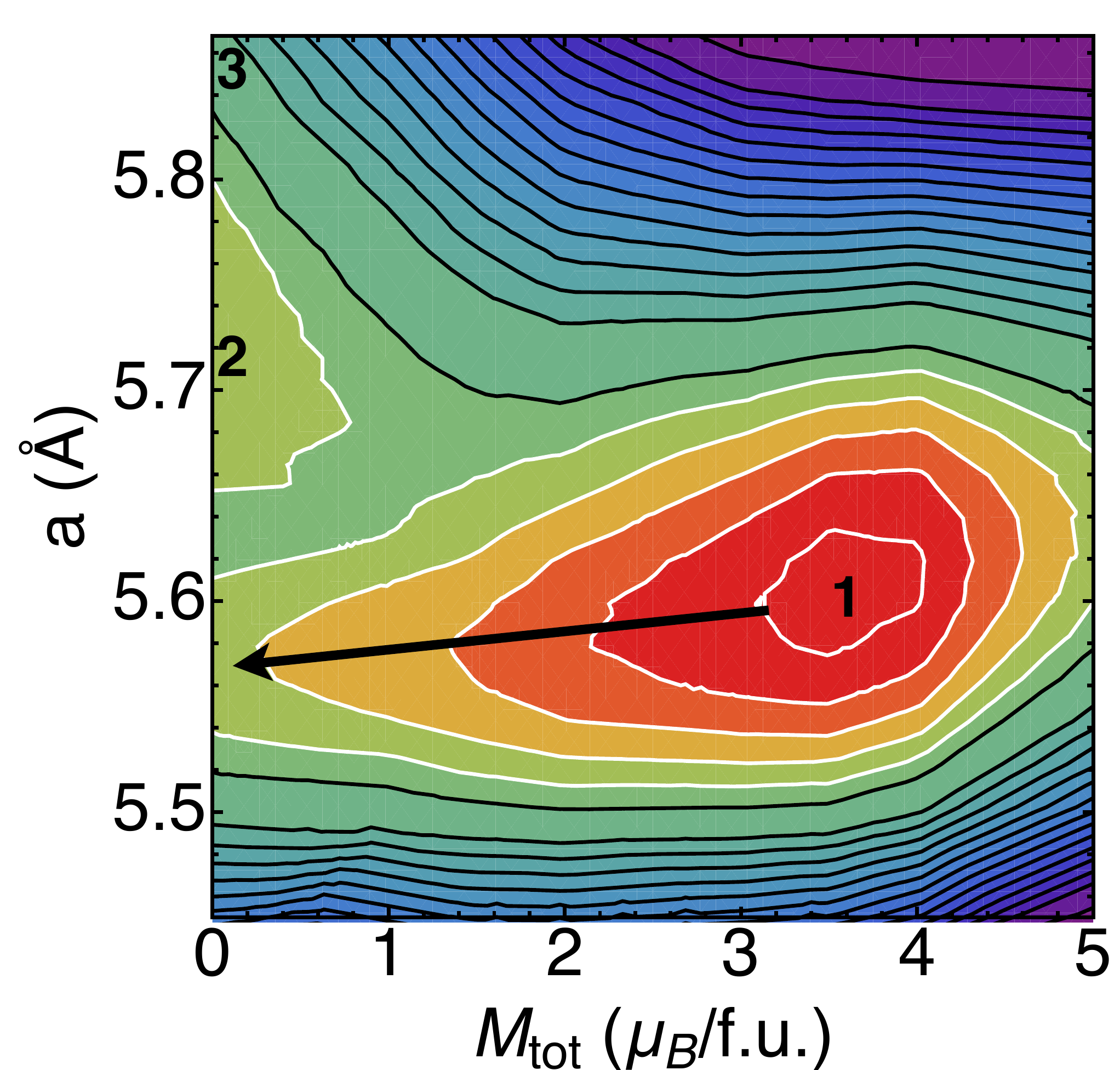}}
\caption{Exemplary energy landscape of pure Mn dependent as function of lattice constant ($a$) and total magnetization (M$_{tot}$) for  spin configuration AF1$_2$, see Fig.~\ref{fig:EM}~(d).
 "2", and "3" mark the AF low and high spin states found at M$_{tot}=0$. For larger values of M$_{tot}$ the spin configuration approaches the FM state, cf.\ Fig.~\ref{fig:EM}~(a).
 The global energy minimum is the FM low spin state "1". Energy contours are given in steps of 20~meV/f.u. relative to state "1".
  \label{fig:EMV}}\end{figure}

As there is a strong magneto-volume coupling in Mn, the magnetic energy landscape changes considerably, if the lattice constant is optimized for the different values of M$_{tot}$ as illustrated for the AF1$_2$ phase in Fig.~\ref{fig:EMV}. 
At M$_{tot}=0$, a systematic decrease of the Mn spins can be found under compression. The system gains 70~meV/f.u. in the relaxed low-spin state ("2" in Fig.~\ref{fig:EMV}).
Globally, the  FM low spin-state ("1" in Fig.~\ref{fig:EMV})  is  most favorable  and the energy landscape is rather flat with respect to reduction of the Mn moment in the FM phase (see arrow in Fig.~\ref{fig:EMV}).

In summary, already pure Mn shows various metastable spin configurations, with local low-spin moments (small volume) or high-spin moments (large volumes).
While the FM low-spin state is the global minimum for all lattice constants, the AF high-spin configurations are more favorable if the system is constrained to the lattice constant of Ni$_7$CoMn$_7$Sn .

\begin{figure*}
\centerline{
\includegraphics[height=0.55\textwidth,angle=270,clip,trim=.5cm 1cm 1cm 1cm]{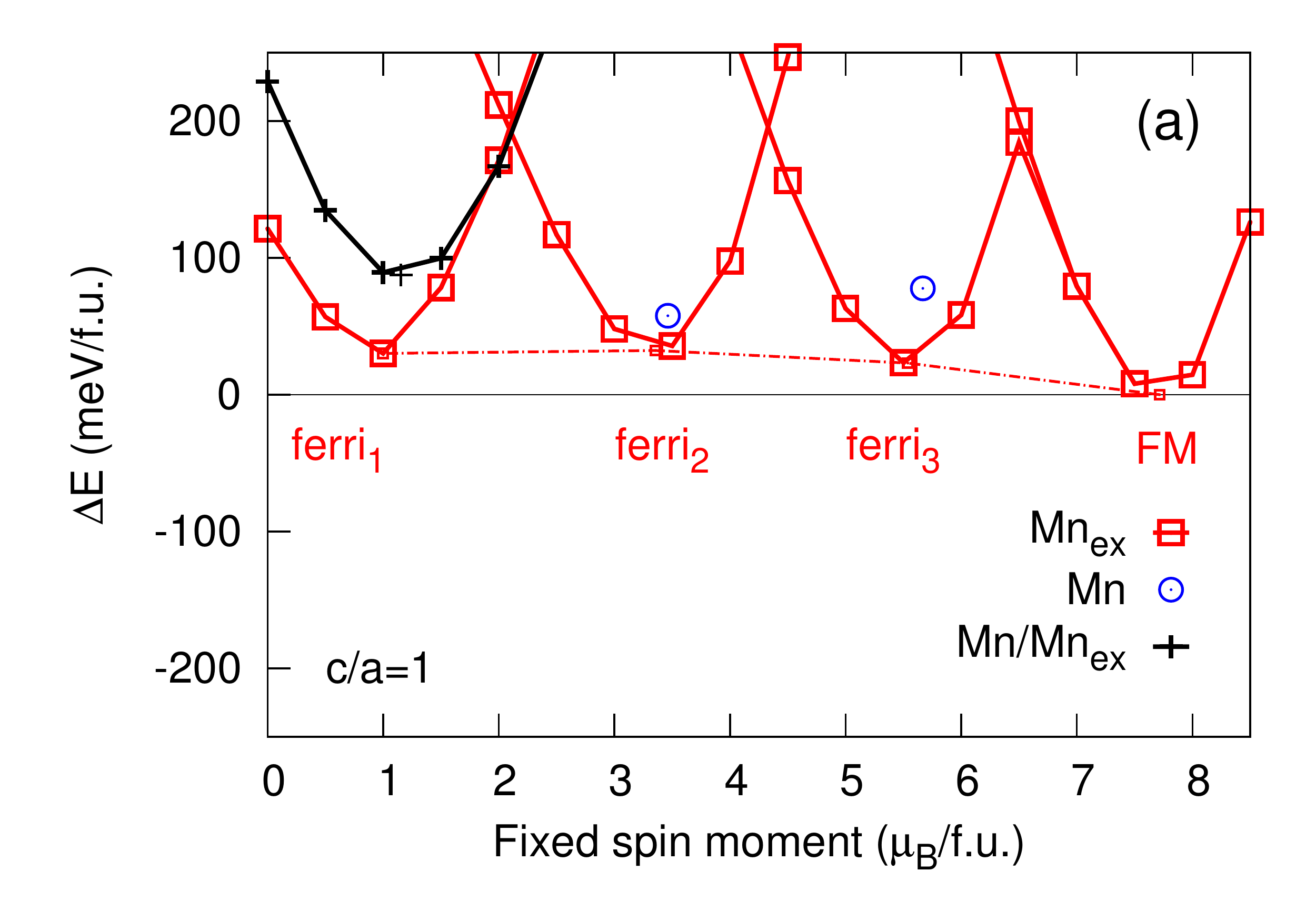}
\includegraphics[height=0.55\textwidth,angle=270,clip, trim=.5cm 1cm 1cm 1cm]{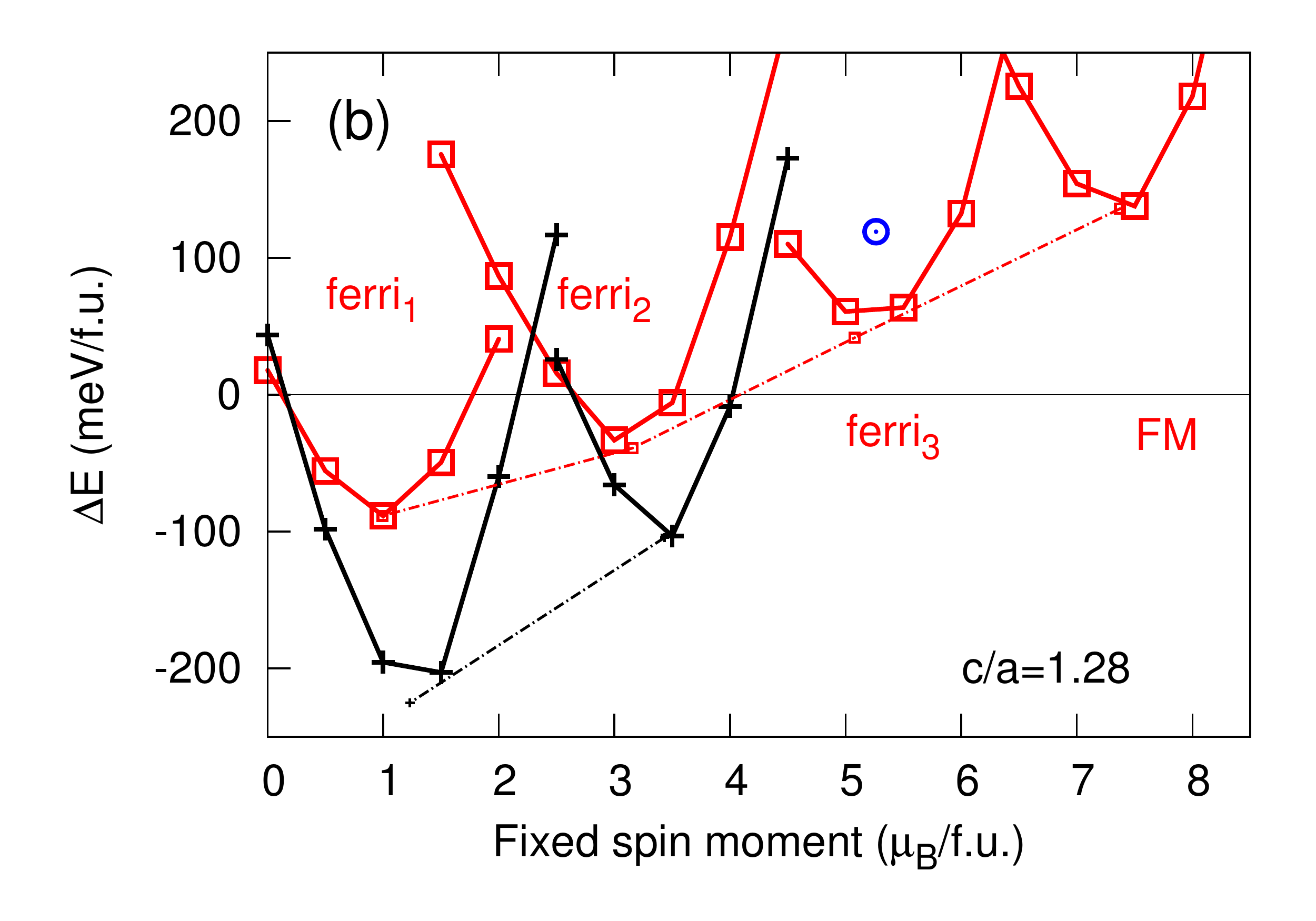}}
\caption{ \label{fig:Energie_ca} Energy landscape of Ni$_7$CoMn$_7$Sn for fully relaxed atomic positions but the lattice constant of the ferri$_1$ phase  for  (a) $c/a=1$ and (b) $c/a=1.28$; 
Solid red lines with open squares: Magnetic states with different numbers of Mn$_{{ex}}^{\downarrow}$ spins.
Black lines with crosses: mixed states with Mn$^{\downarrow}$ and Mn$_{ex}^{\downarrow}$ spins.
Blue circles:  Mn$^{\downarrow}$ spins.
Dotted lines are a guide to the eye only and link relaxed magnetic states without constraints on M$_{tot}$.
}
\end{figure*}

\subsection{The magnetic energy landscape of Ni$_7$CoMn$_7$Sn}
The influence of the lattice constant on the energy landscape of Ni$_7$CoMn$_7$Sn is less pronounced as Ni, Co, and especially the non-magnetic Sn atoms do not show any magneto-volume effect. In contrast to pure Mn, we do not find favorable low-spin states at modified lattice constants and the influence of the magnetic states on the lattice constant is reduced.
For example, for the ferri$_1$ phase, relaxation of the lattice constant for 
 M$_{tot}$ between 0 and 10~$\mu_B$/f.u results in an energy gain in the range of 0.3~meV/f.u., while the change in the lattice constant is below 0.1\%, only. 
The largest changes can be found for the cubic FM phase (the lattice constant increases by 0.4\%, local Mn moments increase by 0.2~$\mu_B$/atom, and the energy is reduced by 6~meV/f.u.) and for the tetragonal mix$_1$ phase (the lattice constant ist reduced by 0.6\%, local Mn moments decrease by 0.5~$\mu_B$/atom, and the energy is reduced by 8~meV/f.u.). However, neither of the energy landscape nor the trends of the magnetic states discussed in the following are modified, cf. also Table~\ref{tab:free}.

\begin{figure}
\centerline{
\includegraphics[height=0.35\textwidth,clip, trim=3cm 3cm 1cm 2.1cm]{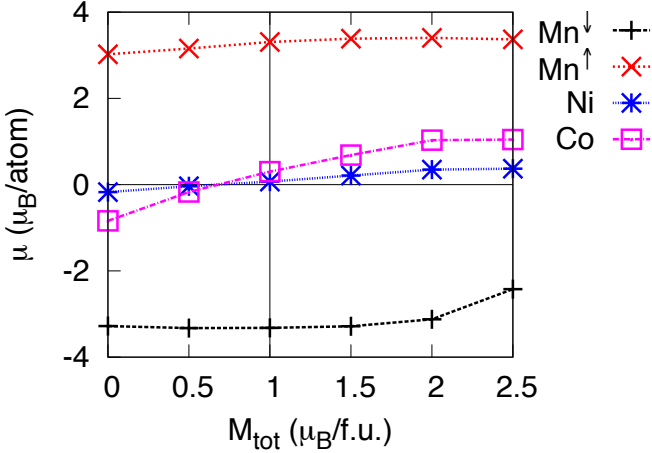}
}
\caption{Change of spin moments in Ni$_7$CoMn$_7$Sn for constraint values of M$_{tot}$ for the mix$_1$ configuration and $c/a=1.28$.\label{fig:moment}}
\end{figure}

So far, FM and ferri$_1$ phases have been discussed in literature \cite{Intermag14} and the additional mix$_{1}$ phase found by MC simulations for $c/a=1.28$ has been discussed in Sec.~\ref{sec:mag}. Other than for pure Mn, see previous section, the FM state is energetically most favorable for $c/a=1$ due to the FM (Ni,Co)-Mn interactions.
Notably, there are many other (meta-)stable magnetic states.
Figure~\ref{fig:Energie_ca} illustrates the energy landscape under constrained values of M$_{tot}$ for generic configurations, compare Table~\ref{tab:free}.
Uniformly, there is a local energy minimum  if all Mn atoms are in the local high-spin state with different possible combinations of Mn$^{\uparrow}$ and Mn$^{\downarrow}$ spins. Apart from the local energy minima, the constraint on $M_{tot}$ enforces reduced Mn spins or Ni and Co moments which are not favorable in the Mn environment, see  Fig.~\ref{fig:moment},
resulting in a steep energy increase, see Fig.~\ref{fig:Energie_ca}.
 In the  chosen supercell containing 7 Mn atoms, the FM state and three different local minima with reduced total moments exist.
Besides the reversal of individual Mn$_{ex}$ spins (ferri$_i$ states), also the reversal of Mn spins on the Mn sublattice (Mn$_i$ states), or a combination of both (mix$_i$ states), result in local energy minima  approximately at the same values of M$_{tot}$, see Fig.~\ref{fig:Energie_ca}.

Surprisingly, the energy minima for the reversal of single Mn$_{ex}$ spins in the ferri$_i$ states and the FM solution are rather similar in energy, see  Tab.~\ref{tab:free}, and the 
  ferri$_{i}$ states are only approximately 30-40~meV/f.u.\ less favorable than the FM state.  
Within our static $T=0$~K simulations we cannot simulate the actual transition between different magnetic states found at finite temperatures. Strictly speaking, only the local energy minima are relevant quantities.
An upper limit for the energy barrier of the transitions can be estimated by the cross points of the static  E($M_{tot}$) curves which are in the range of 200~meV/f.u., only.
In a rough estimate ($E=k_B T$) this corresponds to a thermal activation energy for a spin-flip transition of about 550~K.

Simple combinatorics leads us to expect $mod\left(\frac{N+1}{2}\right)$ local minima of the energy landscape, with $N$ the number of Mn atoms in the cell. 
Indeed,  test simulations with a doubled supercell, see (*) in Table~\ref{tab:free}, result in additional minima with comparable total energies.
Thus, an energy continuum with vanishing energy barriers for thermally or magnetic-field induced  spin flips and a  mixture of different magnetic phases is likely at ambient temperatures without external field. 
 However, rather small magnetic fields are sufficient in order to align all spins in the FM phase with high magnetization. This is a unique feature of Ni-Mn-Sn based Heusler alloys, cf.\ the discussion on Ni$_{50}$Mn$_x$Sn$_{50-x}$ in Ref.~\cite{Behler}.
Also, the different metastable magnetic phases may contribute to the large thermal hysteresis and the irreversible effects, or they may act as nucleation sides for the martensitic transformation.

\begin{figure}
\centerline{
\includegraphics[height=0.35\textwidth,clip,trim=0.5cm 2cm 7.7cm 2cm]{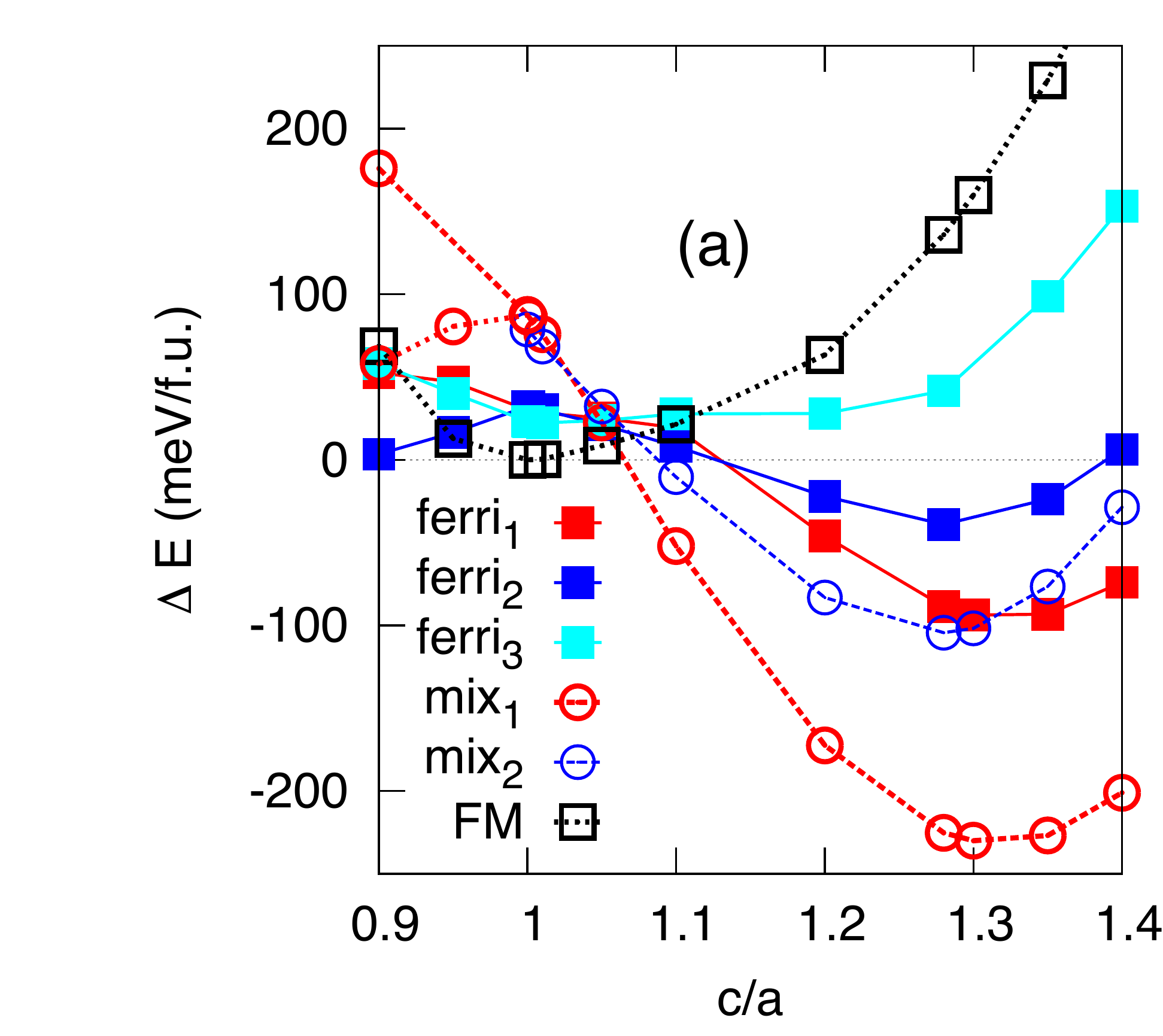}
\includegraphics[height=0.35\textwidth,clip,trim=3cm 2.cm 14cm 2cm]{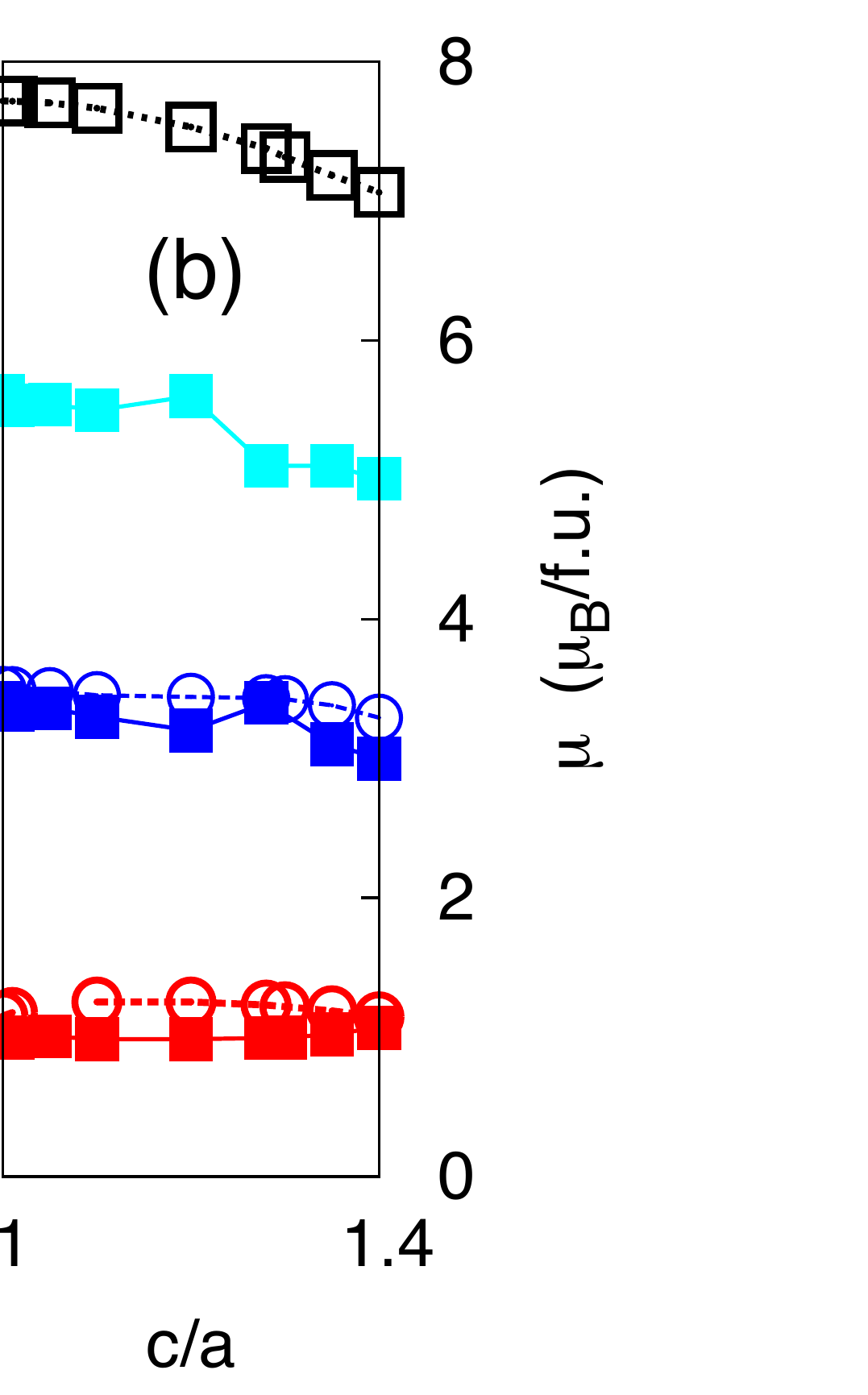}}
\caption{(a) Energy difference between various magnetic configurations as illustrated by symbols, cf.\ text, under tetragonal distortion. (b) Corresponding magnetic moments. For the mix$_1$ phase the stacking of the AF spins along and perpendicular to the tetragonal axis is shown for $c/a<1$. \label{fig:Eca}}
\end{figure}

Figure~\ref{fig:Eca} illustrates the coupling between tetragonal distortion, magnetic moments, and total energy for different (meta-) stable states.
The magnetic moments in the different configurations are rather insensitive to the tetragonal distortion, i.e.\ even the largest change found for the FM phase is   0.7~$\mu_B$/f.u.
In contrast, the relative energies of the different spin configurations change drastically with the tetragonal ratio.
The largest energy change is found for the mix$_1$ configuration. While this phase is  88~meV/f.u.\  higher in energy than the FM state for $c/a=1$, it is the global energy minimum for $c/a\sim1.3$ and already for $c/a>1.06$ this magnetic configuration is most favorable.
 We note that the finite magnetization found is an artifact of the small simulation cell and the disregard of  non-collinear spin alignments, cf.\ Sec.~\ref{sec:mag}, and thus an AF ordering has to be expected for small tetragonal distortions.

For $c/a=1.3$, the FM state is least favorable with an energy difference of  about 390~meV/f.u.\ compared to the mix$_1$ state.
Thus, for a fixed tetragonal distortion, the AF state would be stable up to high magnetic fields. However, the transition between the AF and FM state is related to the structural phase transition. If we let the system relax in our simulations, the FM state always relaxes to the cubic structural state.
This structural transition reduces the energy difference between both magnetic phases by a factor of nearly  two.
As the magnetic energy is vaguely connected with the squared field strength, the structural phase transition considerably reduces the magnetic field strength needed to induce the FM phase.
We note  that the actual energy needed in order to switch between both magnetic states is drastically overestimated by simulations at $T=0$~K.
For example, it has been shown experimentally, that the field needed  to induce the magneto-structural transition in Ni-Mn-In increases by a factor of 4 if the system is cooled down to 4~K \cite{Mejia}. In addition, we do not account for non-collinear magnetic states or magnetic excitations, which may further reduce the magnetic field strength.

In summary, there exists a rich energy landscape of different magnetic states in Ni$_7$CoMn$_7$Sn which are connected by Mn spin flips. For a bulk-like sample a continuum of different such states has to be expected for the cubic phase. 
Smallest tetragonal distortions may destabilize the FM state and the AF state with $c/a\sim 1.3$ is most  favorable. 

\section{Conclusions and Outlook}
\label{sec:conclude}

We have investigated the magnetic energy landscape of (Ni,\ Co)-Mn-Sn (Ni$_7$CoMn$_7$Sn) by means density functional theory simulations using the fixed spin moment approach. To gain further insight in the complex magnetic structure the {\it{ab initio}} results have been mapped on a Heisenberg model and which has been analyzed by Monte Carlo simulations. From this we find a complicated AF spin structure for the low-temperature tetragonal state with its Ne\'el temperature well above the martensitic transition temperature. This spin structure is also most favorable in the {\it{ab initio}} simulations and can be related to the oscillating magnetic exchange interactions between Mn atoms in the tetragonal lattice.
 Similar oscillating magnetic exchange interactions have been found for various Ni-Mn based Heusler alloys in literature. Thus, the appearance of similar complex AF structures may be a general feature and should be taken into account in future investigations.

For the cubic phase we found a transition between the paramagnetic and an ordered magnetic phase above the martensitic transition temperatures and thus a jump in the magnetization of the material and an inverse MCE are likely.
In addition, we have shown that the cubic structure possesses a rich energy landscape with infinite many local energy minima for different spin alignments.
The energy for the flip of single Mn spins is small in the Mn-rich frustrated system.
Already pure Mn shows similar intrinsic properties for the constraint volume. However, in pure Mn, relaxation of the lattice constant stabilizes Mn-low spin phases, instead.
In the Heusler alloy, the Ni, Co, and Sn atoms stabilize a lattice constant which is too large for stable low-spin states. In addition, the Mn-Ni and Mn-Co interaction induces additional FM interactions to the system allowing for a FM phase with high magnetization for the chosen alloy composition.
The dense energy landscape may explain various experimental results. Even for the fixed ordering and composition different magnetic phases are of similar energy and under slight variations of the stoichiometry or maybe thermal fluctuations, large changes in the overall magnetization are possible.
In Ref.~\cite{Fukuda} Fukuda et al. associated the time-dependence of the martensitic transformation in Ni$_{45}$Co$_{5}$Mn$_{36.5}$In$_{13.5}$ with thermally activated cluster formation as nuclei of martensite
with cluster sizes of  (5~nm)$^3$. These clusters or more precisely their formation, can be thought of as being triggered by the different magnetic configurations of the Mn-rich Heusler alloy. The present calculation gives a hint of such magnetic cluster configurations. 
Although magnetic clusters could in principle be simulated in large supercells  this would be very time consuming calculations which we leave for further studies.

\section*{Acknowledgments}
Financial support was granted by the Deutsche Forschungsgemeinschaft (SPP 1599).  H.C.H. acknowledges STandUP for Energy (Sweden) for financial support. The authors thank the Center of Computational Science and Simulation (CCSS) of the University of Duisburg-Essen for computing time. 
\appendix
\section{}
\begin{figure}[b]
\centerline{
\includegraphics[height=0.5\textwidth,angle=270,clip, trim=.5cm 1cm 1cm 1cm]{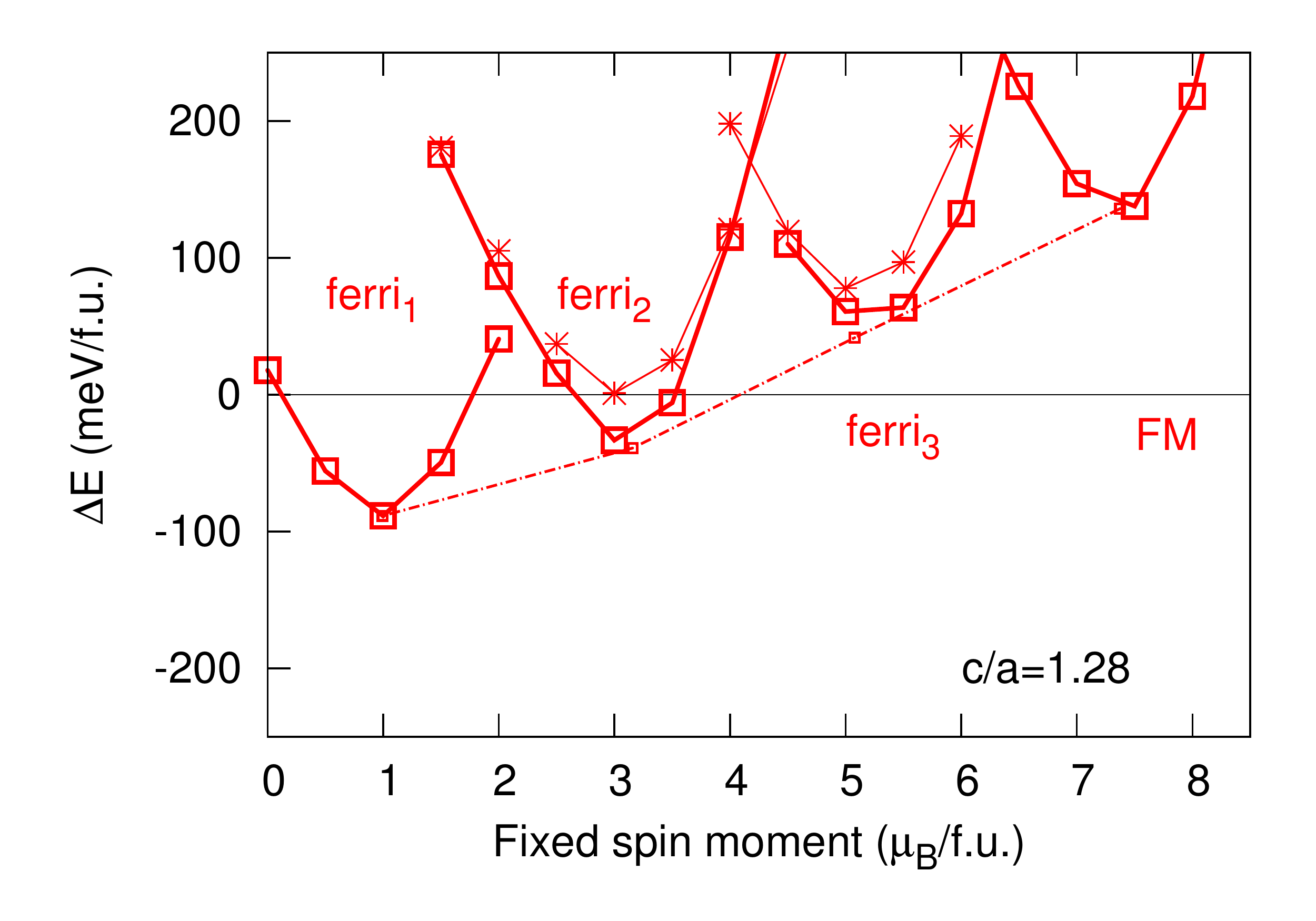}}
\caption{ \label{fig:Efull} Energy landscape for different  numbers of reversed Mn$_{ex}$ spins for 
for c/a=1.28; 
The energy for the reversal of the Mn spins depends on the orientation relative to the tetragonal axis. Open squares: FM Mn$_{ex}$ in  $c$ plane, 
crosses: AF Mn$_{ex}$ in the c plane. The dotted line connects the energies found after the constraint on M$_{tot}$ has been released.
}
\end{figure}
The following section is devoted to technical details. Throughout this paper the supercell shown in Fig.~\ref{fig:struct} has been used to simulate Ni$_7$CoMn$_7$Sn. 
Due to the distribution of Mn and Co atoms, this cell is not cubic and thus tetragonal distortions along a,b, or c are in principle distinguishable.
In addition, the symmetry of the system is further reduced if individual Mn spins are reversed.
While the energy difference between different realizations of the ferrimagnetic states is negligible (below 0.01~meV/f.u.) for $c/a=1$, the E(M$_{tot}$,$c/a\not=$1) curves split for ferri$_2$, and ferri$_3$, depending on the relative alignment of the magnetic phases and the  tetragonal axis, see thin lines in Fig.~\ref{fig:Efull}.
For both states it is most favorable,  if all Mn$_{ex}$ spins within each $c$ plane are aligned FM due to the  pronounced FM exchange interaction at these Mn distances, see Fig.~\ref{fig:Jij}~(c). The same argument holds for the realization of the mix$_i$ phases.  
This state is least favorable for $c/a<1$ as  the increase/decrease of the Mn-Mn distances in the c-planes and along c is reversed compared to $c/a>1$, cf.\ the discussion on the magnetic interactions in Sec.~\ref{sec:mag}.
Although, the energy of the phase can be reduced if the layers with FM coupled Mn, and Mn$_{ex}$ spins are aligned along the b or c axis, the energy difference with the tetragonal distortion is reduced compared to $c/a>1$.
Throughout the paper, we have taken the most favorable realization of each ferri$_i$ or mix$_i$ state into account for the results based on the supercell with 16 Atoms.
We  note that we did not perform systematic tests of different realizations of the ferri$_i$ phases in the doubled supercell (configurations * in Tab~\ref{tab:free}).

\section*{References} 
\bibliographystyle{iopart-num.bst}

\begin{thebibliography}{10}
\expandafter\ifx\csname url\endcsname\relax
  \def\url#1{{\tt #1}}\fi
\expandafter\ifx\csname urlprefix\endcsname\relax\def\urlprefix{URL }\fi
\providecommand{\eprint}[2][]{\url{#2}}

\bibitem{Pecharsky}
Pecharsky V~K and {Gschneidner, Jr} K~A 1997 {\em J. Alloys Compounds\/} {\bf
  260} 98

\bibitem{Gschneidner}
Gschneidner J~K~A, Pecharsky V~K, Pecharsky A~O, Ivtchenko V and Levin E~M 2000
  {\em J. Alloys Compounds\/} {\bf 303-304} 214

\bibitem{Liu2}
Liu J, Gottschall T, Skokov K~P, Moore J~D and Gutfleisch O 2012 {\em Nat.
  Mat.\/} {\bf 11} 620

\bibitem{Planes}
Planes A, Ma{n}osa and Acet M 2009 {\em J. Phys.: Condens. Matter\/} {\bf 21}
  233201

\bibitem{Buchelnikov}
Buchelnikov V~D, Sokolovskiy V~V, Herper H~C, Ebert H, Gruner M~E, Taskaev S~V,
  Khovaylo V~V, Hucht A, Dannenberg A, Ogura M, Akai H, Acet M and Entel P 2010
  {\em Phys. Rev. B\/} {\bf 81} 094411

\bibitem{Sokolovskiy2}
Sokolovskiy V~V, Buchelnikov V~D, Zagrebin M~A, Entel P, Sahoo S and Ogura M
  2014 {\em Entropy\/} {\bf 16} 4992

\bibitem{Graf}
Graf T, Winterlik J, M{\"u}chler L, Fecher G~H, Felser C and Parkin S~P 2013
  {\em Handbook of Magnetic Materials\/} vol~21 (Elsevier) chap Magnetic
  Heusler Compounds

\bibitem{Klaer}
Klaer P, Herper H~C, Entel P, Niemann R, Schultz L, F{\"a}hler S and Elmers H~J
  2013 {\em Phys. Rev. B\/} {\bf 88} 174414

\bibitem{Fabbrici}
Fabbrici S, Kamarad J, Arnold Z, Casoli F, Paoluzi A, Bolzoni F, Cabassi R,
  Solzi M, Porcari G, Pernechele C and Albertini F 2011 {\em Acta Mat.\/} {\bf
  59} 412

\bibitem{Huang}
Huang L, Cong D~Y, Suo H~L and Wang Y~D 2014 {\em Appl. Phys. Lett.\/} {\bf
  104} 132407

\bibitem{NiMnIn3}
Siewert M, Gruner M~E, Hucht A, Herper H~C, Dannenberg A, Chakrabarti A, Barman
  S~R, Singh N, Arr\`{o}yave and Entel P 2012 {\em Adv. Eng. Mater.\/} {\bf 14}
  530

\bibitem{Krenke}
Krenke T, Acet M, Wassermann E~F, Moya X, Ma{\~n}osa L and Planes A 2005 {\em
  Phys. Rev. B\/} {\bf 72} 014412

\bibitem{Sokolovskiy4}
Sokolovskiy V~V, Entel P, Buchelnikov V~D and Gruner M~E 2015 {\em Phys. Rev.
  B\/} {\bf 91} 220409

\bibitem{Cong}
Cong D, Roth S and Schultz L 2012 {\em Acta Mater.\/} {\bf 60} 5335

\bibitem{Intermag14}
Gr\"unebohm A, Comtesse D, Hucht A, Gruner M~E, Maskovskaya A and Entel P 2014
  {\em IEEE Trans. Magn.\/} {\bf 50} 2506004

\bibitem{Umetsu}
Umetsu R~Y, Sheikh A, Ito W, Ouladdiaf B, Ziebeck K~R~A, Kanomata T and Kainuma
  R 2011 {\em Appl. Phys. Lett.\/} {\bf 98} 042507

\bibitem{Ghosh}
Ghosh A and Mandal K 2013 {\em Journal of Alloys and Compounds\/} {\bf 579} 295

\bibitem{Titov}
Titov I, Acet M, Farle M, González-Alonso D, Mañosa L, Planes A and Krenke T
  2012 {\em J.Appl. Phys.\/} {\bf 112} 073914

\bibitem{Bhatti}
Bhatti K~P, El-Khatib S, Srivastava V, James R~D and Leighton C 2012 {\em Phys.
  Rev. B\/} {\bf 85} 134450

\bibitem{Sokolovskiy}
Sokolovskiy V~V, Buchelnikov V~D, Zagrebin M~A, Entel P, Sahoo S and Ogura M
  2012 {\em Phys. Rev. B\/} {\bf 86} 134418

\bibitem{Das}
Das R, Sarma S, Perumal A and Srinivasan A 2011 {\em J. Appl. Phys.\/} {\bf
  109} 07A901

\bibitem{Kanomata}
Kanomata T, Fukushima K, Nishihara H, Kainuma R, Itoh W, Oikawa K, Ishida K,
  Neumann K and Ziebeck K 2008 {\em Mater. Sci. Forum\/} {\bf 583} 119

\bibitem{Jing}
Jing C, Zhang H~L, Chen J~P, Qiao Y~F, Cao S and Zhang J~C 2009 {\em Eur. Phys.
  J. B\/} {\bf 67} 193

\bibitem{Krenke2}
Krenke T, Duman E, Acet M, Wassermann E~F, Moya X, Ma{\~n}osa L and Planes A
  2005 {\em Nature Mater.\/} {\bf 4} 450

\bibitem{Chen}
Chen F, Tong Y, Lu X, Tian B, Li L and Zheng Y 2012 {\em JMEPEG\/} {\bf 21}
  2509

\bibitem{Ye}
Ye M, Kimura A, Miura Y, Shirai M, Cui Y~T, Shimada K, Namatame H, Taniguchi M,
  Ueda S, Kobayashi K, Kainuma R, Shishido T, Fukushima K and Kanomata T 2010
  {\em Phys. Rev. Lett.\/} {\bf 104} 176401

\bibitem{Krumme}
Krumme B, Auge A, Herper H~C, Opahle I, Klar D, Teichert N, Joly L, Ohresser P,
  Landers J, Kappler J~P, Entel P, H\"utten A and Wende H 2015 {\em Phys. Rev.
  B\/} {\bf 91} 214417

\bibitem{Cakir}
\c{C}ak{\i}r A, Righi L, Albertini F, Acet M and Farle M 2015 {\em Act.
  Mater.\/} {\bf 99} 140

\bibitem{Yuan}
Yuan S, Kuhns P~L, Reyes A~P, Brooks J~S, Hoch M~J~R, Srivastava V, James R~D,
  El-Khatib S and Leighton C 2015 {\em Phys. Rev. B\/} {\bf 91} 214421

\bibitem{Cakir2}
\c{C}ak{\i}r A, Acet M and Farle M 2016 {\em Phys. Rev. B\/} {\bf 93} 094411

\bibitem{Cakir3}
\c{C}ak{\i}r A, Acet M, Farle M and Senyshyn A 2014 {\em J. Appl. Phys.\/} {\bf
  115} 043913

\bibitem{Kouvel}
Kouvel J~S and Hartelius C~C 1962 {\em J. Appl. Phys.\/} {\bf 33} 1343

\bibitem{Haeglund}
H\"aglund J 1993 {\em Phys. Rev. B\/} {\bf 47} 566

\bibitem{Moruzzi}
Moruzzi V~L, Marcus P~M, Schwarz K and Mohn P 1986 {\em Phys. Rev. B\/} {\bf
  34} 1784

\bibitem{Duschanek}
Duschanek H, Mohn P and Schwarz K 1989 {\em Physica B\/} {\bf 161} 139

\bibitem{Entel4}
Entel P, Gruner M~E, Adeagbo W~A and Zayak A~T 2008 {\em Mat. Sci. Eng.:A\/}
  {\bf 481} 258

\bibitem{Kresse3}
Kresse G and Joubert D 1999 {\em Phys. Rev. B\/} {\bf 59} 1758

\bibitem{PBE}
Perdew J~P, Burke K and Ernzerhof M 1996 {\em Phys. Rev. Lett.\/} {\bf 77} 3865

\bibitem{Monkhorst}
Monkhorst H~J and Pack J~D 1976 {\em Phys. Rev. B\/} {\bf 13} 5188

\bibitem{fsm}
Schwarz K and Mohn P 1984 {\em J. Phys. F: Met. Phys.\/} {\bf 14} L129

\bibitem{Ogura}
Ogura M, Takahashi C and Akai H 2007 {\em J. Phys.: Condens. Matter\/} {\bf 19}
  365226

\bibitem{Lichtenstein}
Liechtenstein A~I, Katsnelson M~I, Antropov V~P and Gubanov V~A 1987 {\em J.
  Magn. Magn. Mater.\/} {\bf 67} 65

\bibitem{PW91}
Perdew J~P and Wang Y 1992 {\em Phys. Rev. B\/} {\bf 45} 13244

\bibitem{Oedogan}
\"Ozdo\u{g}an K, \ifmmode \mbox{\c{S}}\else \c{S}\fi{}a\ifmmode
  \mbox{\c{s}}\else \c{s}\fi{}\ifmmode \imath \else \i
  \fi{}o\ifmmode~\breve{g}\else \u{g}\fi{}lu E and Galanakis I 2008 {\em J.
  Appl. Phys.\/} {\bf 103} 023503

\bibitem{Gruner}
Gruner M, F\"ahler S and Entel P 2014 {\em Phys. Status Solidi B\/} {\bf 251}
  2067

\bibitem{Kaufmann}
Kaufmann S, R\"o\ss{}ler U~K, Heczko O, Wuttig M, Buschbeck J, Schultz L and
  F\"ahler S 2010 {\em Phys. Rev. Lett.\/} {\bf 104} 145702

\bibitem{Behler}
Behler A, Teichert N, Butta B, Waske A, Hickel T, Auge A, H\"utten A and Eckert
  J 2013 {\em AIP Advances\/} {\bf 3} 122112

\bibitem{Pramanick}
Pramanick S, Chattopadhyay S, Giri S, Majumdar S and Chatterjee S 2014 {\em J.
  Appl. Phys.\/} {\bf 116} 083910

\bibitem{Sasioglu}
\ifmmode \mbox{\c{S}}\else \c{S}\fi{}a\ifmmode \mbox{\c{s}}\else
  \c{s}\fi{}\ifmmode \imath \else \i \fi{}o\ifmmode~\breve{g}\else \u{g}\fi{}lu
  E, Sandratskii L~M and Bruno P 2004 {\em Phys. Rev. B\/} {\bf 70} 024427

\bibitem{Kurtulus}
Kurtulus Y, Dronskowski R, Samolyuk G~D and Antropov V~P 2005 {\em Phys. Rev.
  B\/} {\bf 71} 014425

\bibitem{Rusz}
Rusz J, Bergqvist L, Kudrnovsk\'y J and Turek I 2006 {\em Phys. Rev. B\/} {\bf
  73} 214412

\bibitem{Meinert2}
Meinert M, Schmalhorst J~M and Reiss G 2011 {\em Journal of Physics: Condensed
  Matter\/} {\bf 23} 116005

\bibitem{Sasioglu3}
\ifmmode \mbox{\c{S}}\else \c{S}\fi{}a\ifmmode \mbox{\c{s}}\else
  \c{s}\fi{}\ifmmode \imath \else \i \fi{}o\ifmmode~\breve{g}\else \u{g}\fi{}lu
  E, Sandratskii L~M and Bruno P 2008 {\em Phys. Rev. B\/} {\bf 77} 064417

\bibitem{Entel5}
Entel P, Siewert M, Gruner M~E, Herper H~C, Comtesse D, Arr\'oyave R, Singh N,
  Talapatra A, Sokolovsiky V~V, Buchelnikov V~D and Albertini F 2013 {\em Eur.
  Phys. J. B\/} {\bf 86} 65

\bibitem{Sliwko}
Sliwko V, Mohn P and Schwarz K 1994 {\em J. Phys. Cond. Mat.\/} {\bf 6} 6557

\bibitem{Hafner}
Hafner J and Spi\ifmmode~\check{s}\else \v{s}\fi{}\'ak D 2005 {\em Phys. Rev.
  B\/} {\bf 72} 144420

\bibitem{Uebayashi}
Uebayashi K, Shimizu H and Yamada H 2006 {\em Mater. Trans.\/} {\bf 47} 456

\bibitem{Mejia}
Mej{\i}a C~S, Zavareh M~G, Nayak A~K, Skourski Y, Wosnitza J, Felser C and
  Nicklas M 2015 {\em J. Appl. Phys\/} {\bf 117} 17E710

\bibitem{Fukuda}
Fukuda T, Kakeshita T and h~Lee Y 2014 {\em Acta Mater.\/} {\bf 81} 121

\end{thebibliography}
\providecommand{\newblock}{}

\end{document}